\documentclass[amssymb,aps,twocolumn,showpacs, superscriptaddress,nofootinbib]{revtex4}
\usepackage[]{graphicx}
\usepackage[]{amsmath}
\usepackage{amsthm}
\usepackage{hyperref}
\usepackage{extraipa}

\newcommand{\be}{\begin{equation}}
\newcommand{\ee}{\end{equation}}

\def\cB{\mathcal{B}}

\def\cH{\mathcal{H}}

\def\cM{\mathcal{M}}
\def\cN{\mathcal{N}}

\def\Tr{\mathrm{Tr}}

\newtheorem{theorem}{Theorem}
\newtheorem{lemma}{Lemma}
\def\eq#1{Eq.~\eqref{eq:#1}}
\def\fig#1{Fig.~\ref{fig:#1}}

\begin{document}

\title{Markov entropy decomposition: a variational dual for quantum belief propagation}
\author{David Poulin}
\email{David.Poulin@USherbrooke.ca}
\affiliation{D\'epartement de Physique, Universit\'e de Sherbrooke, Qu\'ebec, J1K 2R1, Canada}
\author{Matthew B. Hastings}
\affiliation{Department of Physics, Duke University, Durham, NC 27708}
\affiliation{Microsoft Research, Station Q, Elings Hall, University of California, Santa Barbara, CA 93106, USA.}

\date{\today}

\begin{abstract}
We present a lower bound for the free energy of a quantum many-body system at finite temperature. This lower bound is expressed as a convex optimization problem with linear constraints, and is derived using strong subadditivity of von Neumann entropy and a relaxation of the consistency condition of local density operators. The dual to this minimization problem leads to a set of quantum belief propagation equations, thus providing a firm theoretical foundation to that approach. The minimization problem is numerically tractable, and we find good agreement with quantum Monte Carlo for the spin-$\frac 12$ Heisenberg anti-ferromagnet in two dimensions.  This lower bound complements other variational upper bounds.  We discuss applications to Hamiltonian complexity theory and give a generalization of the structure theorem of \cite{HJPW03a} to trees in an appendix.
\end{abstract}


\maketitle

Describing the properties of a local quantum system is perhaps the central problem of theoretical physics.  However, the exponential growth of the Hilbert space with system size makes it prohibitive to even write down the state of a system with even a modest number of sites.  For this reason, variational methods, such as
matrix product states used in DMRG
\cite{OR95a,DMNS98a,VPC04b,Vid05a} and their higher dimensional generalizations\cite{VC04a,SDV06a}, 
 are a central tool, describing a state with a small number of parameters, allowing a practical optimization of the energy.

All these methods provide an upper bound to the free energy and the quality of the approximation cannot be assessed directly.
In this Letter, we present a {\em lower} bound to the free energy that nicely complements variational approaches. We use strong subadditivity (SSA) of von Neumann entropy \cite{LR73a} to approximate the system's entropy by a local quantity. This approximation is exact when the system is a Markov network \cite{LP08a}---i.e., when its long-range correlations arise due to correlations over shorter distances---but in general provides a lower bound to the true entropy. By relaxing the consistency constraints on the reduced density operators of the systems, we find a formula for the free energy expressed as a convex minimization problem with linear constraints. 

Our formula for the free energy is similar to the Bethe free energy \cite{B35a}---and its generalization by Kikuchi \cite{Kik51a}---, but differs by a crucial ordering of the lattice sites. This distinction is responsible for the lower bound obtained by our method, in contrast to Bethe's and Kikuchi's approximations which are uncontrolled. The dual of the minimization problem provides a set of quantum belief propagation equations similar to those presented in \cite{Has07b,LP08a,PB08a}. This connection provides a solid theoretical foundation to understand the success and limitations of quantum belief propagation. Similar connections \cite{Yed01a} and algorithms \cite{GJ07a} have been found in the classical setting.

\noindent{\it Markov entropy decomposition---}Consider a lattice of $N$ spins that we label from $1$ to $N$. The labeling
of the sites chosen will determine the order in which we apply our procedure later.  The Hamiltonian of the system is a sum of geometrically local terms $H = \sum_X h_X$ where $X$ labels subsets of $\{1,\ldots N\}$ and locality means that $h_X = 0$ when the radius of $X$ is larger than some constant $w$. Given the density matrix $\rho$ of the system, we can compute the average energy $E(\rho) = \Tr(\rho H) = \sum_X \Tr (\rho_X h_X)$ from knowledge of only the reduced density matrices  $\rho_X \equiv \Tr_{\overline X} \rho$ on small local regions, that can be obtained from the partial trace of $\rho$ over the complement $\overline X$ of $X$.

At finite temperature $T$, we are interested in the system's free energy $F(T)\equiv \min_\rho \{E(\rho)-TS(\rho)\}$.  Unlike the energy, the entropy $S(\rho) \equiv \Tr(\rho\log\rho)$ cannot be evaluated in general from knowledge of only the reduced density matrices $\rho_X$ over regions $X$ of finite radius.   We define an approximate way of doing this evaluation.
For every site $k$, define a subset of sites $\cN_k$ consisting of ``neighboring" sites. There is no unique prescription for the choice of $\cN_k$, but it is useful to imagine that they consist of a set of sites located within a finite distance from $k$. With trivial manipulations, we can rewrite the entropy of the system in the form of an ``entropy chain rule" $S(\rho) = \sum_{k=1}^N S(k | \{<\! k\})$ where the conditional entropy of a region $X$ given region $Y$ is $S(X|Y) \equiv S(X\cup Y) - S(Y)$, the entropy of any region $X$ is denoted $S(X) \equiv S(\rho_X) = -\Tr(\rho_X \log \rho_X)$, and we use the notation $\{<\! k\} = \{1,2,\ldots, k-1\}$.

\begin{figure}
\includegraphics[width=7cm]{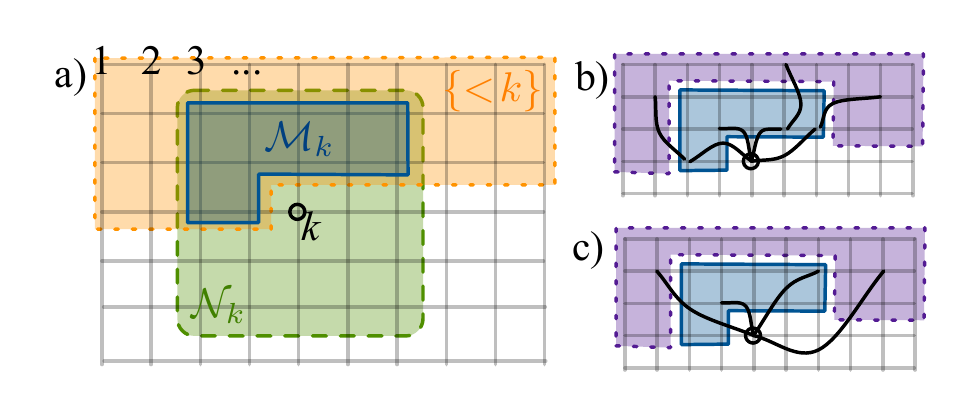}
\caption{(Color online) a) The Marrkov shield (shown in blue) $\cM_k$ is the intersection of the neighborhood (green) of $k$ and the sites preceding $k$ (orange). b) The entanglement (represented by black lines) between site $k$ and the preceding sites is all mediated by the Markov shield: the state of the first $k$ sites can be constructed by adding one extra spin to the state of the first $k-1$ site and coupling it only to the sites of the shield \cite{HJPW03a}. This turns inequality \eq{SSA} into an equality. c) There is direct entanglement between site $k$ and the sites preceding $k$, so the Markov entropy is not equal to the true entropy, but it is an upper bound.}
\label{fig:Markov_shield}
\end{figure}

Quantum entropy $S$ obeys SSA\cite{LR73a}, which implies the bound
\begin{equation}
S(k|\{<\! k\}) \leq S(k | \{<\! k\} \cap \cN_k)=S(k|\cM_k),
\label{eq:SSA}
\end{equation}
where we define $\cM_k= \{<\! k\} \cap \cN_k$.  We call $\cM_k$ the ``Markov shield" of site $k$, see \fig{Markov_shield}.
We can define the Markov entropy $S_M(\rho) \equiv \sum_{k=1}^N S(k | \cM_k)$ which upper bounds the system's entropy. Because each term in that sum can be computed from the reduced density matrices on site $k$ and its Markov shield, the Markov entropy, unlike the entropy $S$, is suitable for direct numerical calculations.

Returning to the free energy calculation, we now have the bound $F(T) \geq  F_M(T) \equiv \min_{\rho} \{E(\rho) - T S_M(\rho)\}$. The Markov free energy $F_M$ of any given state is equal to its true free energy if SSA is saturated with the given choice of Markov shields as shown in \fig{Markov_shield}. Because both $E$ and $S_M$ can be evaluated from the density matrix of constant-size regions $X$, we can express $F_M(\rho) = E(\rho)-T S_M(\rho)$ as a function of some set of reduced density operators $\{\rho_X\}$ and write $ F_M(T) = \min_{\{\rho_X\} \in \Omega} F_M(\{\rho_X\})$, where $\Omega$ denotes the set of {\em consistent} reduced density matrices that are all obtainable from some global density matrix $\rho$, i.e. $\Omega \equiv \big\{ \{\rho_X\} : \exists \rho, \ \rho_X = \Tr_{\overline X} \rho ,\ \forall X \big\}$. 

Unfortunately, verifying consistency of a set of reduced density matrices $\{\rho_X\}$ is a difficult problem, it is QMA-complete \cite{L06a}, so it is very unlikely that $\Omega$ can be characterized efficiently. Thus, we will make one more approximation and enlarge the set $\Omega$ to the set $\tilde \Omega$ of all {\it locally} consistent reduced density matrices that agree on overlapping regions, i.e. $\tilde \Omega \equiv \big\{ \{\rho_X\} : \Tr_{\overline{X\cap Y}} \rho_X = \Tr_{\overline{X\cap Y}} \rho_Y,\ \forall (X,Y)\big\}$. Since all reduced density matrices in $\Omega$ are derived from one global $\rho$, it should be clear that $\Omega \subset \tilde \Omega$, and as a consequence
\begin{align}
F_{\rm MED}(T) &\equiv \min_{\{\rho_X\} \in \tilde \Omega} F_M(\{\rho_X\}) \leq F_M(T) \leq F(T).
\label{eq:MED}
\end{align}

Equation~\eqref{eq:MED} defines our numerical method that we call the Markov entropy decomposition (MED) scheme. The Markov free energy $F_M(\{\rho_X\})$ to be minimized to evaluate $F_{\rm MED}(T)$ is a convex function\footnote{That conditional entropy is convex also follows from SSA.} over the cone of semi-positive operators $\{\rho_X\}$ subject to some linear constraints specified in the definition of $\tilde\Omega$. Thus, it is suitable for numerical optimization.

\begin{figure}
\includegraphics[width=7cm]{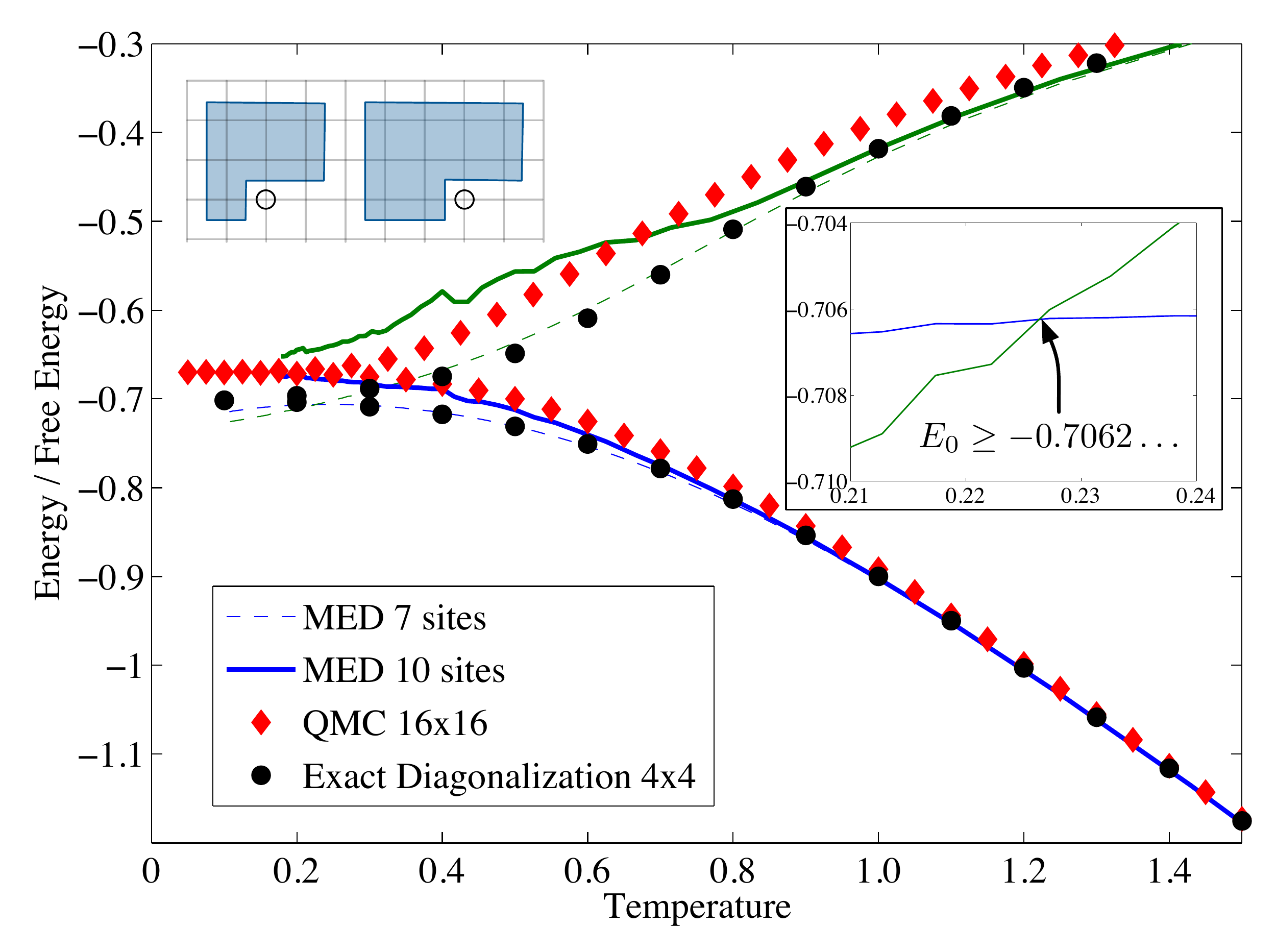}
\caption{Numerical results obtained from MED for the spin-$\frac 12$ Heisenberg antiferromagnet on a 2D square lattice. The energy (green) and free energy (blue) are obtained for a 7- and 10-site Markov shield, of shape illustrated in the upper left corner. Results  are compared to exact diagonalization of a $4\times 4$ lattice and quantum Monte Carlo. The crossing of energy and free energy curve (negative entropy, 7-site shield) provides a lower bound to the ground energy. }
\label{fig:results}
\end{figure}

\noindent{\it Numerical results on translationally invariant systems---}The procedure simplifies greatly when applied to translationally invariant systems.  If we assume that all density matrices $\rho_X$ are related by translational symmetry, the Markov free energy is a function of a single density matrix. 
We have numerically investigated this method with a spin-$\frac 12$ antiferromagnetic Heisenberg model on an infinite  two-dimensional square lattice. We have used a Markov shield of size 7 and 10, so that the main computational task of our program was exact diagonalization of (non-sparse) matrices of size $2^8$ and $2^{11}$ respectively. Figure \ref{fig:results} compares our results to other methods.

The MED free energy with the 10-site shield is in excellent agreement with quantum Monte Carlo for the entire temperature range. This agreement with QMC is better than the one obtained from exact diagonalization (ED) of a $4\times 4$ lattice.  In fact, those diagonalization results are very well approximated by MED with a 7-site shield. Here we see the biggest advantage of MED: because of the constraints imposed on the minimization, the results converge to the thermodynamic limit faster than ED.

Since entropy is positive and $\partial F/\partial T = -S$,  we see that the free energy is a monotonically decreasing function of temperature. However, the Markov entropy $S_M(\{\rho_X\})$ can be negative when the global consistency is not satisfied, and we indeed observe that the slope of the free energy changes sign near $T=0.2$. Markov entropy becomes negative where $F_M(T)=E_M(T)$. Since $F_M(T) \leq E(\rho(0)) - TS_M(\rho(0)) \leq E_0$, the crossing point of the MED energy and free energy obtained with the 7-site shield gives a lower bound $E_0\geq -0.7062...$ to the true ground state energy of the system. 

We have used this technique to lower bound the ground state energy of the one-dimensional model. Results obtained with a $k$-site neighborhood are in good agreement with ED results on a chain of length roughly $2k$ (with periodic boundary conditions). This can be understood from the fact that the ground-state entropy of a block of $\ell$ sites, $S(\ell)$, is an increasing function of $\ell$ for $\ell \leq k$, and then decreases to reach 0 when $\ell = 2k$ since the entire system is in a pure state. Thus, enforcing a positive Markov entropy density $S_M = S(k) - S(k-1)$ compels the system in our simulations to behave as it were on a lattice of size $2k$, even though we are manipulating states of $k$ spins, providing some heuristic explanation for the improved convergence, compared to ED, seen above.

All these lower bounds on the ground state energy and the lower bounds on the free energy, would be rigorous if the convex optimization problem were solved exactly. However, all our results are subject to numerical error. We used fairly elementary minimization methods (conjugate gradient) and more elaborate techniques that exploit the special features of this problem are likely to improve the results; we hope that this Letter will stimulate research in this direction. Numerical fluctuations are most prominent in the energy, while the free energy curve is rather smooth. 
The fluctuations are largest near the specific heat peak; to understand this, consider the free energy $E-TS_M$ as a function of $E$, assuming
for simplicity that $S_M$ equals the correct entropy $S(E)$.  At a minimum of $F$, $\frac{\partial F}{\partial E}=0$, and
$\frac{\partial^2 F}{\partial E^2} =\frac 1{Tc}$ and so for large $c$, the basin around the minimum
is shallow, increasing numerical error.
We now describe an alternate approach, a dual problem, which connects to quantum belief propagation.  If this dual problem could be turned into a {\it variational} dual problem (a concave function whose maximum equals the minimum of the Markov free energy), it would provide mathematically rigourous lower bounds on $F$.

\begin{figure}
\includegraphics[width=5cm]{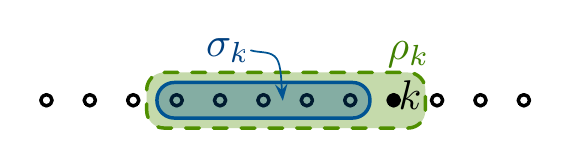}
\vspace*{-0.5cm}
\caption{The density matrix $\rho_k$ describes the state of sites $k-n$ to $k$ while the states $\sigma_k$ is for sites $k-n$ to $k-1$, with $n=5$ in this example. }
\label{fig:chain}
\end{figure}

\noindent{\it Dual problem: quantum belief propagation---}Consider a length-$N$ spin chain and define density matrices $\rho_k$ and $\sigma_k$ associated to segments $k-n$ to $k$ and $k-n$ to $k-1$ respectively, as in \fig{chain}. In this case, the minimization problem defined at \eq{MED} becomes
\begin{align}
\sum_{k=n}^{N} \Big( & \Tr\{\rho_k[\hat H_k +\log \rho_k - I\otimes A_k - B_{k}\otimes I +\mu_k]\} \nonumber \\
&- \Tr\{\sigma_k [\log\sigma_k - A_{k-1} -B_k +\nu_k]\}\Big)
\end{align}
where for $k=n,\ldots N$, the matrices $A_k$ and $B_k$ and the scalars $\mu_k$ and $\nu_k$ are Lagrange multipliers used to enforce $\Tr_1\rho_k = \sigma_{k+1}$, $\Tr_n\rho_k = \sigma_{k}$, and the trace normalization of $\rho_k$ and $\sigma_k$ respectively, and $A_{N}=0$.  Above, $\hat H_k$ is the part of the Hamiltonian supported on sites $k$ to $k+n$ properly weighted to avoid double counting, and we have set temperature $T=1$ to avoid cluttering equations. Taking derivatives with respect to $\rho_k$ and $\sigma_k$ yields
\begin{align}
\hat H_k + \log\rho_k -I A_k -B_{k} I +\mu'_k = 0 \\
\log \sigma_k -A_{k-1} - B_k +\nu'_k = 0
\end{align}
where $\nu_k' = \nu_k+1$ and $\mu_k' = \mu_k+1$, and we have dropped the $\otimes$ symbols. These equations, together with the constraints imposed on the reduced density matrices, give a set of self-consistent mean-field equations
\begin{align}
%A_k &= \log(\Tr_n \rho_{k+1}) +\bar A_{k+1}-\nu_{k+1} \label{eq:MF} \nonumber \\
%&= \log(\Tr_n e^{-\hat H_{k+1}+IA_{k+1} +\bar A_{k}I -\mu'_{k+1}}) +\bar A_{k+1}-\nu_{k+1} \nonumber \\
A_{k-1} &= \log(\Tr_n \rho_{k}) -B_{k}+\nu'_{k} \nonumber \\
&= \log(\Tr_n e^{-\hat H_{k}+IA_{k} +B_kI -\mu'_{k}}) -B_k+\nu'_k \label{eq:MF1}\\
B_{k+1} &= \log(\Tr_1\rho_k) -A_k +\nu'_{k+1} \nonumber \\
&= \log(\Tr_1 e^{-\hat H_{k}+IA_{k} +B_{k}I -\mu'_{k}}) -A_k+\nu'_{k+1} . \label{eq:MF2}
\end{align}
One can show that any solution to these equations is a minimum of the Markov free energy \eq{MED}. Because this function is convex, the solution to Eqs.~(\ref{eq:MF1},\ref{eq:MF2}) is unique.
% up the the global symmetry $A_k \rightarrow A_k+ICI$, $B_k \rightarrow B_k - ICI$ where $C$ is any operator on $n-2$ sites.

We can conceive an iterative procedure to approach solutions to Eqs.~(\ref{eq:MF1},\ref{eq:MF2}). Starting from an initial guess for the $A_k$ and $B_k$, we obtain new guesses by inserting these values into Eqs.~(\ref{eq:MF1},\ref{eq:MF2}) which provides new values, and recurse. Renaming $A_{k-1} = \log m_{k\rightarrow k-1}$ and $B_{k+1} = \log m_{k\rightarrow k+1}$, we recognize Eqs.~(\ref{eq:MF1},\ref{eq:MF2}) as almost the {\em belief propagation} prescription of \cite{LP08a,PB08a}
\begin{align}
m_{k\rightarrow k-1} &\propto \Tr_n(\Lambda_k\odot m_{k+1\rightarrow k} \odot m_{k-1\rightarrow k}) \odot m_{k-1\rightarrow k}^{-1}\label{eq:m1}\\
m_{k\rightarrow k+1} &\propto \Tr_1(\Lambda_{k}\odot m_{k+1\rightarrow k} \odot m_{k-1\rightarrow k}) \odot m_{k+1\rightarrow k}^{-1}\label{eq:m2}\\
%\sigma_k &\propto m_{k-1\rightarrow k}\odot m_{k+1\rightarrow k} \label{eq:b1}\\
\rho_k &\propto \Lambda_k\odot m_{k+1\rightarrow k}\odot m_{k-1\rightarrow k} \label{eq:b2}
\end{align}
where all proportionality constants can be set by normalization and $\Lambda_k = \exp(-\hat H_{k})$. The $\odot$ product is defined by $A\odot B = \exp(\log A + \log B)$. We note a subtle difference between these belief propagation equations and those of \cite{LP08a,PB08a}. If the action of the partial trace and the $\odot$ product were commutative as they are in the classical case, the two appearances of the term $m_{k-1\rightarrow k}$ in \eq{m1} would cancel, and similarly for  $m_{k+1\rightarrow k}$ in \eq{m2}. These cancellations were assumed in \cite{LP08a,PB08a}, based on heuristic arguments and numerical evidences. However, we see that they are required to establish a direct connection with the MED. Any fixed point of the iteration equations for messages $m$ yields a lower bound to the free-energy of the system. Moreover, as in \cite{Has07b,LP08a,PB08a}, this iterative procedure can be used to evaluate other quantities such as correlation functions.

\noindent{\it State reconstruction and probabilistically checkable proofs (PCP)---} Given a global quantum state $\rho$, such that SSA is saturated for the given choice of Markov shields, we can reconstruct the global state from the local state.  Using the structure theorem of \cite{HJPW03a}, we have
$\log(\rho_{\{<\! k+1\} })=\log(\rho_{\{<\! k\}})+\log(\rho_{k\cup \cM_k} -\log(\rho_{\cM_k})$.  Iterating this procedure allows us to reconstruct the global state from the local state. In the Appendix, we extend this idea and show that any state saturating SSA on a tree graph is the thermal state of a Hamiltonian that is the sum of local, commuting terms. This procedure may help address the structure of topologically ordered states, since many lattice models with topological order saturate SSA with an appropriate choice of shields\cite{LW05a} (see the Appendix).

Deciding whether the ground state energy of a classical Hamiltonian on $N$ particles is 0 or greater than $N\epsilon$ for some positive constant $\epsilon$ is a very difficult problem. In general, it is NP-complete, by the famous PCP theorem\cite{ALMS98a}.  The analogous decision problem for a quantum Hamiltonian \cite{AALV09a} is in QMA \cite{K02a}, but it is not known to be QMA-complete (this is the quantum PCP conjecture).
While this question concerns zero temperature, it is equivalent to determining whether the free energy becomes negative at temperature $T < \epsilon/\log d$ where $d$ is the number of levels of each particle.
It is easy to verify if a set of operators $\{A_k,B_k\}$ are a solution to  Eqs.~(\ref{eq:MF1},\ref{eq:MF2}), so the problem of lower bounding the free energy of a quantum system using the Markov entropy decomposition is in NP. Thus, one way to disprove the quantum PCP conjecture would be to find a rigorous upper bound to this lower bound, e.g., by analyzing its scaling as a function of the size of the Markov shield.  State reconstruction may prove useful here.

\noindent{\it Multi-patch MED---}We now discuss a possible extension of our method. Let $F_M^1$ and $F_M^2$ denote the Markov free energy formulas obtained from two different of neighborhoods in our procedure. Clearly, $F_M^{\rm max} = \max_{k} F_M^k$ is a  lower bound to the free energy. The convex function
\begin{equation*}
F_{\rm MED}^{1,2}(T) \equiv \min_{\{\rho_X\} \in \tilde \Omega} \max_k F_M^k(\{\rho_X\})
\end{equation*}
is an even better lower bound. That is, instead of minimizing $F_M^1$ and $F_M^2$ separately, we minimize their maximum, subject to the constraint that the reduced density matrices used to compute the two formulas are locally consistent with one another.

\begin{figure}
\includegraphics[width=5.2cm]{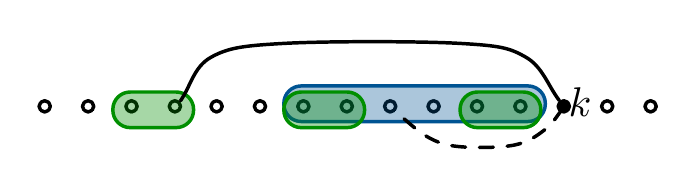}
\vspace*{-0.5cm}
\caption{The blue and green regions are two different Markov shields for site $k$ (the green neighborhood is not a connected region). }
\label{fig:chainMP}
\end{figure}

In particular, the shapes of $\cM_1$ and $\cM_2$ can be chosen to capture correlations on different length scales of the system. Figure~\ref{fig:chainMP} illustrates two such choices. The blue region captures the short-scale entanglement (depicted by a dashed line) while the green neighborhood captures the long-range entanglement (full line). The free energy formula obtained by the combination of both regions is forced to assign reduced density matrices compatible with both type of correlations.

\noindent{\it Discussion---} MED is on the one hand a possible  numerical tool for studying the thermodynamics of quantum systems in
a more accurate way than is possible using exact diagonalization.  On the other hand, it provides a
theoretical basis for the quantum belief propagation procedure developed previously to study disordered quantum systems; while we focused in translationally invariant systems above, we can apply the procedure more generally, e.g. to quantum spin glasses \cite{KU97b}, treating each reduced density matrix $\rho_X$ as an independent variable.  Finally, it offers a physics-inspired procedure that may help tackle outstanding problems in quantum computational complexity.

\noindent{\it Acknowledgments---}We thank Roger Melko for providing the QMC data. Computational resources were provided by RQCHP. D.P. receives financial support from  NSERC and FQRNT.

%\bibliographystyle{/Users/dpoulin/archive/hsiam}
%\bibliographystyle{/Users/dpoulin/archive/qubib}
%\bibliography{/Users/dpoulin/archive/qubib}

\begin{thebibliography}{21}
\expandafter\ifx\csname natexlab\endcsname\relax\def\natexlab#1{#1}\fi
\expandafter\ifx\csname bibnamefont\endcsname\relax
  \def\bibnamefont#1{#1}\fi
\expandafter\ifx\csname bibfnamefont\endcsname\relax
  \def\bibfnamefont#1{#1}\fi
\expandafter\ifx\csname citenamefont\endcsname\relax
  \def\citenamefont#1{#1}\fi
\expandafter\ifx\csname url\endcsname\relax
  \def\url#1{\texttt{#1}}\fi
\expandafter\ifx\csname urlprefix\endcsname\relax\def\urlprefix{URL }\fi
\providecommand{\bibinfo}[2]{#2}
\providecommand{\eprint}[2][]{\url{#2}}

\bibitem[{\citenamefont{\"{O}stlund and Rommer}(1995)}]{OR95a}
\bibinfo{author}{\bibfnamefont{S.}~\bibnamefont{\"{O}stlund}} \bibnamefont{and}
  \bibinfo{author}{\bibfnamefont{S.}~\bibnamefont{Rommer}},
  \bibinfo{journal}{Phys. Rev. Lett.} \textbf{\bibinfo{volume}{75}},
  \bibinfo{pages}{3537} (\bibinfo{year}{1995}).

\bibitem[{\citenamefont{Dukelsky et~al.}(1998)\citenamefont{Dukelsky,
  Mart{\'i}n-Delgado, Nishino, and Sierra}}]{DMNS98a}
\bibinfo{author}{\bibfnamefont{J.}~\bibnamefont{Dukelsky}},
  \bibinfo{author}{\bibfnamefont{M.}~\bibnamefont{Mart{\'i}n-Delgado}},
  \bibinfo{author}{\bibfnamefont{T.}~\bibnamefont{Nishino}}, \bibnamefont{and}
  \bibinfo{author}{\bibfnamefont{G.}~\bibnamefont{Sierra}},
  \bibinfo{journal}{Europhys. Lett.} \textbf{\bibinfo{volume}{43}},
  \bibinfo{pages}{457} (\bibinfo{year}{1998}).

\bibitem[{\citenamefont{Verstraete et~al.}(2004)\citenamefont{Verstraete,
  Porras, and Cirac}}]{VPC04b}
\bibinfo{author}{\bibfnamefont{F.}~\bibnamefont{Verstraete}},
  \bibinfo{author}{\bibfnamefont{D.}~\bibnamefont{Porras}}, \bibnamefont{and}
  \bibinfo{author}{\bibfnamefont{J.~I.} \bibnamefont{Cirac}},
  \bibinfo{journal}{Phys. Rev. Lett.} \textbf{\bibinfo{volume}{93}},
  \bibinfo{pages}{227205} (\bibinfo{year}{2004}).

\bibitem[{\citenamefont{Vidal}(2007)}]{Vid05a}
\bibinfo{author}{\bibfnamefont{G.}~\bibnamefont{Vidal}},
  \bibinfo{journal}{Phys. Rev. Lett} \textbf{\bibinfo{volume}{99}},
  \bibinfo{pages}{220405} (\bibinfo{year}{2007}), \eprint{cond-mat/0512165}.

\bibitem[{\citenamefont{Verstraete and Cirac}(2004)}]{VC04a}
\bibinfo{author}{\bibfnamefont{F.}~\bibnamefont{Verstraete}} \bibnamefont{and}
  \bibinfo{author}{\bibfnamefont{J.~I.} \bibnamefont{Cirac}},
  \emph{\bibinfo{title}{Renormalization algorithms for quantum-many body
  systems in two and higher dimensions}} (\bibinfo{year}{2004}),
  \eprint{cond-mat/0407066}.

\bibitem[{\citenamefont{Shi et~al.}(2006)\citenamefont{Shi, Duan, and
  Vidal}}]{SDV06a}
\bibinfo{author}{\bibfnamefont{Y.-Y.} \bibnamefont{Shi}},
  \bibinfo{author}{\bibfnamefont{L.-M.} \bibnamefont{Duan}}, \bibnamefont{and}
  \bibinfo{author}{\bibfnamefont{G.}~\bibnamefont{Vidal}},
  \bibinfo{journal}{Phys. Rev. A} \textbf{\bibinfo{volume}{74}},
  \bibinfo{pages}{022320} (\bibinfo{year}{2006}).

\bibitem[{\citenamefont{Lieb and Ruskai}(1973)}]{LR73a}
\bibinfo{author}{\bibfnamefont{E.}~\bibnamefont{Lieb}} \bibnamefont{and}
  \bibinfo{author}{\bibfnamefont{M.}~\bibnamefont{Ruskai}},
  \bibinfo{journal}{J. Math. Phys.} \textbf{\bibinfo{volume}{14}},
  \bibinfo{pages}{1938} (\bibinfo{year}{1973}).

\bibitem[{\citenamefont{Leifer and Poulin}(2008)}]{LP08a}
\bibinfo{author}{\bibfnamefont{M.}~\bibnamefont{Leifer}} \bibnamefont{and}
  \bibinfo{author}{\bibfnamefont{D.}~\bibnamefont{Poulin}},
  \bibinfo{journal}{Ann. Phys.} \textbf{\bibinfo{volume}{323}},
  \bibinfo{pages}{1899} (\bibinfo{year}{2008}).

\bibitem[{\citenamefont{Bethe}(1935)}]{B35a}
\bibinfo{author}{\bibfnamefont{H.}~\bibnamefont{Bethe}},
  \bibinfo{journal}{Proc. Roy. Sco. A} \textbf{\bibinfo{volume}{150}},
  \bibinfo{pages}{552} (\bibinfo{year}{1935}).

\bibitem[{\citenamefont{Kikuchi}(1951)}]{Kik51a}
\bibinfo{author}{\bibfnamefont{R.}~\bibnamefont{Kikuchi}},
  \bibinfo{journal}{Phys. Rev.} \textbf{\bibinfo{volume}{81}},
  \bibinfo{pages}{988} (\bibinfo{year}{1951}).

\bibitem[{\citenamefont{Hastings}(2007)}]{Has07b}
\bibinfo{author}{\bibfnamefont{M.~B.} \bibnamefont{Hastings}},
  \bibinfo{journal}{Phys. Rev. B} \textbf{\bibinfo{volume}{76}},
  \bibinfo{pages}{201102(R)} (\bibinfo{year}{2007}).

\bibitem[{\citenamefont{Poulin and Bilgin}(2008)}]{PB08a}
\bibinfo{author}{\bibfnamefont{D.}~\bibnamefont{Poulin}} \bibnamefont{and}
  \bibinfo{author}{\bibfnamefont{E.}~\bibnamefont{Bilgin}},
  \bibinfo{journal}{Phys. Rev. A} \textbf{\bibinfo{volume}{77}},
  \bibinfo{pages}{052318} (\bibinfo{year}{2008}).

\bibitem[{\citenamefont{Yedidia}(2001)}]{Yed01a}
\bibinfo{author}{\bibfnamefont{J.~S.} \bibnamefont{Yedidia}}, in
  \emph{\bibinfo{title}{Advanced mean field methods: theory and practice}}
  (\bibinfo{publisher}{MIT Press}, \bibinfo{year}{2001}),
  p.~\bibinfo{pages}{21}.

\bibitem[{\citenamefont{Globerson and Jaakkola}(2007)}]{GJ07a}
\bibinfo{author}{\bibfnamefont{A.}~\bibnamefont{Globerson}} \bibnamefont{and}
  \bibinfo{author}{\bibfnamefont{T.}~\bibnamefont{Jaakkola}}, in
  \emph{\bibinfo{booktitle}{11th Int. Conf. on Art. Intel. and Stat.}}
  (\bibinfo{year}{2007}).

\bibitem[{\citenamefont{Liu}(2006)}]{L06a}
\bibinfo{author}{\bibfnamefont{Y.-K.} \bibnamefont{Liu}},
  \bibinfo{journal}{Proc. RANDOM} p. \bibinfo{pages}{438}
  (\bibinfo{year}{2006}).

\bibitem[{\citenamefont{Hayden et~al.}(2004)\citenamefont{Hayden, Jozsa, Petz,
  and Winter}}]{HJPW03a}
\bibinfo{author}{\bibfnamefont{P.}~\bibnamefont{Hayden}},
  \bibinfo{author}{\bibfnamefont{R.}~\bibnamefont{Jozsa}},
  \bibinfo{author}{\bibfnamefont{D.}~\bibnamefont{Petz}}, \bibnamefont{and}
  \bibinfo{author}{\bibfnamefont{A.}~\bibnamefont{Winter}},
  \bibinfo{journal}{Comm. Math. Phys.} \textbf{\bibinfo{volume}{246}},
  \bibinfo{pages}{359} (\bibinfo{year}{2004}).

\bibitem[{\citenamefont{Levin and Wen}(2006)}]{LW05a}
\bibinfo{author}{\bibfnamefont{M.}~\bibnamefont{Levin}} \bibnamefont{and}
  \bibinfo{author}{\bibfnamefont{X.-G.} \bibnamefont{Wen}},
  \bibinfo{journal}{Phys. Rev. Lett.} \textbf{\bibinfo{volume}{96}},
  \bibinfo{pages}{110405} (\bibinfo{year}{2006}).

\bibitem[{\citenamefont{Arora et~al.}(1998)\citenamefont{Arora, Lund, Motwani,
  and Szegedy}}]{ALMS98a}
\bibinfo{author}{\bibfnamefont{S.}~\bibnamefont{Arora}},
  \bibinfo{author}{\bibfnamefont{C.}~\bibnamefont{Lund}},
  \bibinfo{author}{\bibfnamefont{R.}~\bibnamefont{Motwani}}, \bibnamefont{and}
  \bibinfo{author}{\bibfnamefont{M.}~\bibnamefont{Szegedy}},
  \bibinfo{journal}{J. ACM} \textbf{\bibinfo{volume}{45}}, \bibinfo{pages}{501}
  (\bibinfo{year}{1998}).

\bibitem[{\citenamefont{Aharonov et~al.}(2009)\citenamefont{Aharonov, Arad,
  Landau, and Varizani}}]{AALV09a}
\bibinfo{author}{\bibfnamefont{D.}~\bibnamefont{Aharonov}},
  \bibinfo{author}{\bibfnamefont{I.}~\bibnamefont{Arad}},
  \bibinfo{author}{\bibfnamefont{Z.}~\bibnamefont{Landau}}, \bibnamefont{and}
  \bibinfo{author}{\bibfnamefont{U.}~\bibnamefont{Varizani}}, in
  \emph{\bibinfo{booktitle}{Proc. of the 41st Ann. ACM Symp. on Theo. of
  Comp.}} (\bibinfo{year}{2009}), p. \bibinfo{pages}{417}.
  
  \bibitem[{\citenamefont{Kitaev}(2002)}]{K02a}
\bibinfo{author}{\bibfnamefont{A.~Y.} \bibnamefont{Kitaev}}, in
  \emph{\bibinfo{booktitle}{Classical and quantum computation}}, edited by
  \bibinfo{editor}{\bibfnamefont{A.~Y.} \bibnamefont{Kitaev}},
  \bibinfo{editor}{\bibfnamefont{A.~H.} \bibnamefont{Shen}}, \bibnamefont{and}
  \bibinfo{editor}{\bibfnamefont{M.~N.} \bibnamefont{Vyalyi}}
  (\bibinfo{year}{2002}).

\bibitem[{\citenamefont{Kope{\'c} and Usadel}(1997)}]{KU97b}
\bibinfo{author}{\bibfnamefont{K.}~\bibnamefont{Kope{\'c}}} \bibnamefont{and}
  \bibinfo{author}{\bibfnamefont{K.~D.} \bibnamefont{Usadel}},
  \bibinfo{journal}{Phys. Rev. Lett.} \textbf{\bibinfo{volume}{78}},
  \bibinfo{pages}{1988} (\bibinfo{year}{1997}).

\end{thebibliography}

\appendix
\section{Structure of Thermal States Saturating Strong Subadditivity Locally}
\begin{figure}
\includegraphics[width=5cm]{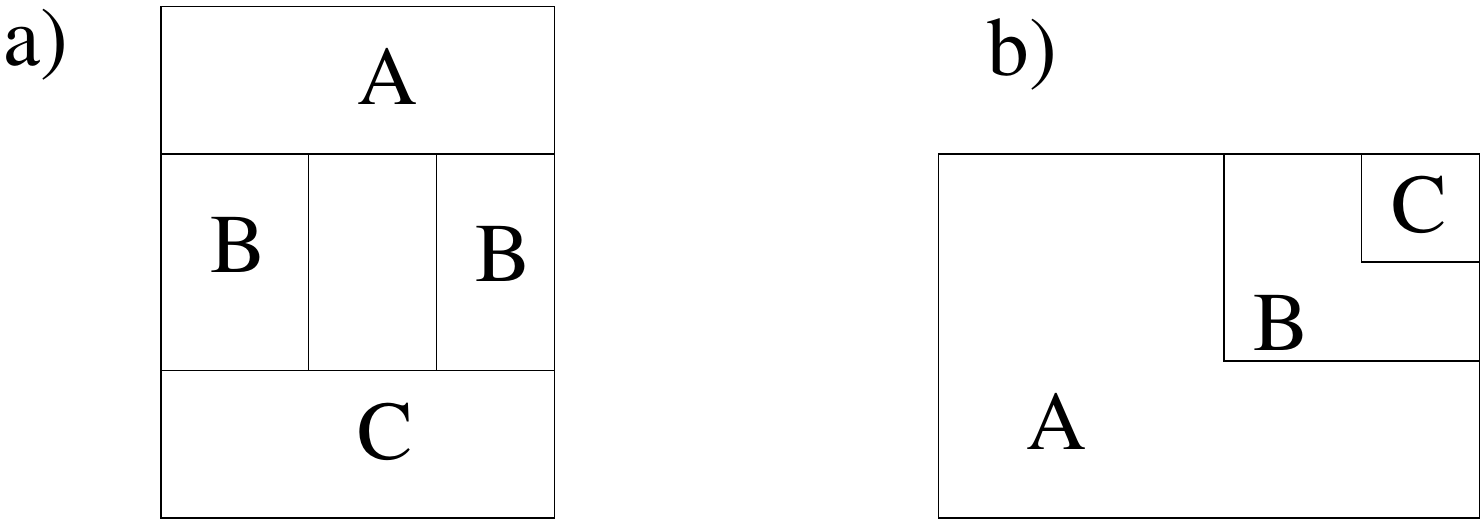}
\caption{a)Topological entanglement entropy from a configuration with topology change. b)No topology change, and strong subadditivity is saturated.}
\label{fig:topo}
\end{figure}

In this appendix we discuss the structure of states which exactly saturate strong subadditivity.  Our major result is
a statement about the structure of such states on a tree graph.  However, first we would like to briefly discuss
why such states occur even in the {\it ground state} of finite dimensional topologically ordered lattice models.  Consider a lattice model such as
those considered in \cite{LW05a} which has a vanishing correlation length but also has topological order.  Such a model displays an
interesting correction to the entropy, called ``topological entanglement entropy".  This causes the entropy of a region to have a term
which is proportional to the boundary of the region, plus a constant which depends upon the topology of the region.  In \cite{LW05a}, this
constant is extracted by considering a sum and difference of entropies over different regions.  However, this sum and difference
exactly corresponds to a conditional mutual information of three regions $A,B,C$, for a particular choice of the regions.  If we pick it
so that $ABC$ is an annulus, as shown in Fig.~(\ref{fig:topo})a.  Then, the mutual information between $A$ and $C$ conditioned on $B$ is proportional to the entanglement entropy term that Levin and Wen consider, and strong subadditivity can be used to determine the sign of
the topological correction to the entanglement entropy.  

So, from this we learn that for certain choices of sites and shields in such
a model we will {\it not} see a saturation of the conditional mutual information.  Indeed, problems occur whenever there is a topology change.
However, we can instead consider a case as in Fig.~(\ref{fig:topo})b in which all three regions are contractable.  In this case, the
conditional mutual information vanishes in these models and strong subadditivity is saturated.  Thus, we can in many cases find
a sequence of sites to add and a choice of shields such that strong subadditivity is saturated.  In particular, let us consider a system on a sphere.  Then if we choose the set $\{1,...,k\}$ to be contractible at every step but the last (this is why we chose the sphere), and the neighborhoods are chosen to be a small circle around each site, then strong subadditivity will be saturated at every stage.  One can verify that when the last site is added, strong subadditivity is saturated also.

Since strong subadditivity is saturated at each stage, this enables us to write the projector onto the ground state of the system
as a matrix product operator with bounded bond dimension.  To do this, we iterate the result that saturation of strong subadditivity
implies that $\rho_{ABC}=\rho_{BC}^{1/2} \rho_B^{-1/2} \rho_{AB} \rho_{B}^{-1/2} \rho_{BC}^{1/2}$; each such operator $\rho_{AB}$ has
bounded bond dimension, and as a result the operator $\rho_{ABC}$ has bounded bond dimension.  There exists some product state
such that $\rho_{ABC}$ acting on that state is non-zero.  Applying $\rho_{ABC}$ to that state then gives a representation of the
ground state as a matrix product state or PEPS (projected entangled pair state) \cite{VC04a}.

After this discussion, we now turn to the structure of states saturating strong subadditivity on a tree, in which case we can prove
much more about the structure of the states.
We will prove that any such density matrix can be written as
$\rho=Z^{-1} \exp(-H)$ for $H$ a sum of commuting local operators and $Z$ a normalization constant.  We prove
this result as a corollary of a result which generalizes the structure
theorem of \cite{HJPW03a} to tree graphs.  Our structure theorem on tree graphs
has the physical interpretation that there are two types of
correlations between nodes on the graph.  There are classical
correlations, which can be long-ranged, but are always mediated by
correlations between intermediate nodes, and there are quantum
correlations which are limited to nearest neighbors.  

We begin with the special case of a one dimensional system,
with the sites $1,2,...,k$ chosen
in order along a line.
For simplicity, assume that strong subadditivity becomes saturated at the shortest nontrivial length scale,
when $\cM_k=\{k-1\}$.
Note that if strong subadditivity is saturated on some larger length scale (for example, if we only saturate strong subadditity when
$\cM_k=\{k-2,k-1\})$, then by a rescaling of the system, grouping several sites into one site, we can reduce 
to the case when $\cM_k=\{k-1\}$.

The structure theorem implies that the Hilbert space $\cH_k$ on any site $k-1$ can be decomposed as a direct sum
\be
\cH_{k-1}=\bigoplus_j \cH_{{k-1}^L(j)} \otimes \cH_{{k-1}^R(j)},
\ee
so that
\be
\rho_{\{1,...,k\}}=\bigoplus_j q_{k-1}(j) \rho_{\{1,...,k-2\},{k-1}^L(j)} \otimes \rho_{{k-1}^R(j),k},
\ee
where the $q_{k-1}(j)$ are a probability distribution.

Let us assume  that the temperature $T=1$, for notational simplicity.  Write the density matrix $\rho=Z^{-1} \exp(-H)$ for some $H$.  We
will show how to write such an $H$ as a sum of local, commuting operators.
Let $P_k(j)$ denote the operator on site $k$ which projects onto $B(k)^L_j \otimes B(k)^R_j$.  Define
\be
q_{k}(j|i)=\Tr(P_k(j) P_{k-1}(i) \rho)/\Tr(P_{k-1}(i) \rho),
\ee
Then,
\be
\rho=\frac 1Z \exp\Big[-\sum_{k=1}^N H_k\Big],
\ee
where
\be
H_1=-\sum_j P_1(j) \ln(q_1(j)],
\ee
and
\be
H_k=
\sum_{i,j} P_k(j) P_{k-1}(i) \Bigl(\ln(q_k(j|i))+\ln(\rho_{{k-1}^R,{k}^L})\Bigr)
\ee
for $k>1$.  The operators  $H_k$ commute for different $k$ due to the tensor product structure of Hilbert spaces $B(k)^L_j \otimes B(k)^R_j$, and they
are local as required.
We omit a proof that this procedure is correct, since it is a special case of our more general result on trees, below.

We now describe a similar procedure which can be applied 
to any tree graph.  First, some definitions.  We define a density matrix $\rho_{ABC}$ to be a Markov chain $A-B-C$ if strong
subadditivity is saturated, so that $S(C|BA)=S(C|B)$.  We define a density matrix on a multi-partite system to be a Markov network if
there is a graph, with each subsystem corresponding to a node of the graph, such that, given
any three disjoint sets $A,B,C$ of nodes of the graph such that all paths
from any node in $A$ to any node in $C$ must past through a node in $B$, the density matrix $\rho_{ABC}$ is a Markov chain on $A-B-C$.
We will later consider tree graphs, with nodes labelled $1,...,N$.
Let node $1$ be called the ``root" of the tree.
For each node other than the root, the ``parent" of that
node is considered to be the neighbor of that which is closer than the root, and the daughters are considered to be the other neighbors of
that graph.  We let $p(i)$ be the parent function: $p(i)$ is the parent node of node $i$ if $i>1$.
Let the nodes be ordered such that if $i<j$ then the path from node $i$ to the root does not pass through node $j$ (i.e., node $i$ is not a daughter, grand-daughter, etc... of node $j$).  
We say that such a tree is a ``Markov tree" if, for each node $k>1$
we have
\be
S(k|\{<\! k\}) = S(k | \{<\! k\} \cap \cN_k)=S(k|\cM_k),
\ee
where the Markov shield $\cN_k$ of $k$ is the parent of node $k$.  Note that a Markov tree is simply a Markov network that is a tree graph; while we have defined Markov trees with a particular choice of root, they would be Markov trees for any choice of the root.

With these definitions, we will prove a result which
extends the quantum Hammersley-Clifford theorem derived in~\cite{LP08a}:
\begin{theorem}
Any Markov network on a tree can be expressed as $\rho = \frac 1Z \exp(-H)$ where $H$ is the sum of local, commuting terms.
\begin{proof}
This is a corollary of theorem (\ref{thm:treestruct}) proven below as the
operators $H_i$ in that theorem are local and commuting.
\end{proof}
\end{theorem}

The following Lemma will be useful in proving theorem (\ref{thm:treestruct}).
\begin{lemma}
Let $\rho_{ABC}$ be a Markov chain on $A-B-C$ and suppose that $P$ is a projector onto a subspace of $\cH_B$ such that $[P,\rho_{ABC}] = 0$. Then $P\rho_{ABC} P$ is also a Markov chain on $A-B-C$.
\label{lem:sum}
\begin{proof}
We can write $\rho_{ABC} = \rho_{AB^1C} \oplus \rho_{AB^2C}$. Saturation of SSA is equivalent \cite{LP08a} to the equality $\log\rho_{ABC} = \log\rho_{AB} + \log\rho_{BC} - \log \rho_{B}$. The proof follows from the fact that $\log(X\oplus Y) = \log X \oplus \log Y$. 
\end{proof}
\end{lemma}

We first prove a special case of our result on a tree, which can be thought of as a generalization of the structure theorem.
Since we will use this terminology later, first make a definition.  Given a multi-partite state $\rho$ on $N$ subsystems, labelled $1,...,N$ and referred to as ``nodes",
define a {\bf splitting of node $k$} to be
a decomposition of the Hilbert space $\cH_k$ on node $k$ as
\be
\cH_k=\bigoplus_j \cH_k(j),
\ee
where each Hilbert space $\cH_k(j)$ can be decomposed into a tensor product
\be
\cH_k(j)=\bigotimes_{i \neq k, 1\leq i \leq N} \cH_{k\rightarrow i}(j),
\ee
such that 
the density matrix $\rho$ can be expressed as
\be
\rho=\bigoplus_j q(j) \bigotimes_{i\neq k, 1 \leq i \leq N} \rho_{\cH_{k\rightarrow i}(j),i},
\label{eq:node}
\ee
where $\rho_{\cH_{k\rightarrow i}(j),i}$ is a density matrix on $\cH_{k\rightarrow i}(j)$ and $i$.

We now prove that
\begin{theorem}
\label{thm:node}
Consider any Markov tree with $N$ nodes, such that all nodes, other than the root, are daughters of the root.  Then, there exists
a splitting of the root.
\begin{proof}
The proof is inductive.  Let node $1$ be the root to simplify notation.
Assume that we have proven the theorem when the Markov tree  has only $N-1$ nodes (the case $N=3$ is the
structure theorem of \cite{HJPW03a}).  Apply the structure theorem with the three subsystems $A=\{2,...,N-1\},B=\{1\},C=\{N\}$ to show that
there exists a decomposition of the Hilbert space $\cH_1$ on node $1$ into
$\cH_1=\bigoplus_j \cB(j)$, where
$\cB(j)=\cB(j)^L \otimes \cB(j)^R$ with
$\rho=\sum_j q(j) \rho_{A,\cB(j)^L} \otimes \rho_{\cB(j)^R,C}$.
By lemma (\ref{lem:sum}), 
$\rho_{A,\cB(j)^L} \otimes \rho_{\cB(j)^R,C}$ is a Markov tree.  Thus, $\rho_{A,\cB(j)^L}$ is a Markov tree
on a graph of $N-1$ nodes.  Thus, applying the inductive assumption, there
exists a decomposition of $\cB(j)^L$ into a direct sum of Hilbert spaces
\be
\cB(j)^L=\bigoplus_k \cH_{1,j}(k),
\ee
where each Hilbert space $\cH_{1,j}(k)$ can be decomposed into a tensor product
\be
\cH_{1,j}(k)=\bigotimes_{i \geq 2}^{N-1} \cH_{(1,j) \rightarrow i}(k),
\ee
such that 
the density matrix $\rho_{A,\cB(j)^L}$ can be expressed as
\be
\rho_{A,\cB(j)^L}=\bigoplus_k r_j(k) \bigotimes_{i\geq 2}^{N-1} \rho_{\cH_{(1,j) \rightarrow i}(k),i},
\ee
for some probability distribution $r_j(k)$.
Then, let
\be
\cH_1((j,k))=\cH_{1,j}(k) \otimes \cB(j)^R,
\ee
so that
\be
\cH_1((j,k))=
\bigotimes_{i \geq 2}^{N-1} \cH_{(1,j) \rightarrow i}(k) \otimes \cB(j)^R.
\ee
Treating the two indices $j,k$ as a single index, this gives a splitting for $N$ sites.
\end{proof}
\end{theorem}

We now prove a structure theorem for trees:
\begin{theorem}
\label{thm:treestruct}
Consider a tree graph with $N$ different nodes, labelled $1,...,N$, forming a Markov tree.
Then, for each node $k$ there exists a decomposition of the Hilbert space $\cH_k$ on that node into a sum of Hilbert spaces
\be
\cH_k=\bigoplus_j \cH_k(j),
\ee
where each Hilbert space $\cH_k(j)$ can be decomposed into a tensor product
\be
\cH_k(j)=\bigotimes_{i} \cH_{k\rightarrow i}(j),
\ee
where the product ranges over nodes $i$ which are neighbors of node $k$,
such that the following properties hold.
We use $P_k(j)$ to denote the projector  onto $\cH_k(j)$, we use $q_k(j)$ to denote $\Tr(\rho P_k(j))$,
and we define
\be
q_{k}(j|i)=\Tr(P_k(j) P_{p(k)}(i) \rho)/\Tr(P_{p(k)}(i) \rho).
\ee
Then,
the density matrix $\rho$ can be expressed as
\be
\label{rhosum}
\rho=\exp\Big[-\sum_{k=1}^N H_k\Big],
\ee
where
\be
H_1=-\sum_j P_1(j) \ln(q_1(j)],
\ee
and
\begin{eqnarray}
H_k= 
&\sum_{i,j} P_k(j) P_{p(k)}(i) 
\Bigl(&\ln(q_k(j|i))+
\\ \nonumber
&&\ln(\rho_{\cH_{p(k)\rightarrow k}(i),\cH_{k \rightarrow p(k)}(j)})\Bigr).
\end{eqnarray}
(Note that in case $q_k(j)=0$ for any $k$, we define $\exp[ln(q_k(j))]=0$ and define conditional probabilities in which $q_k(j)$ appears in the denominator arbitrarily.)
\begin{proof}
For each node $k$, consider the subgraph consisting of $k$ and all of its neighbors.  
Let the decomposition $\cH_k(j)=\bigotimes_{i} \cH_{k\rightarrow i}(j)$ in the statement of this theorem be the
splitting given in the previous theorem for the given subgraph.

For use later, we define a new coarse-grained graph, as follows.
Let $k$ be the root of the new graph, labelled $k$.  For each neighbor $i$ of node $k$ on the original graph, group that neighbor and
all nodes connected to that neighbor by a path that does not go through node $k$ into one node on the new graph, and label
that new node $\tilde i$.  See Fig.~(\ref{fig:Coarse}).  We claim that the splitting above also provides a splitting on the coarse-grained graph.  This holds because the density matrix on the coarse-grained graph can be constructed from the density matrix on the subgraph by applying a super-operator which is a product of super-operators on each of the nodes as follows.
Let $\rho_{k,\{i\in n(k)\}}$ be the density matrix on $k$ and its neighbors $i$, tensored with the identity on the remaining nodes.
Let $n(k)$ be the set of neighbors of $k$.  We have 
\be
\rho_{k,\{\tilde i\}}=
\Bigl( \prod_{i\in n(k)} 
\rho_{\tilde i}^{1/2} \rho_{i}^{-1/2} \Bigr)
\rho_{k,\{i\in n(k)\}}
\Bigl( \prod_{i\in n(k)} 
\rho_{i}^{-1/2} 
\rho_{\tilde i}^{1/2} \Bigr)
\ee
 by strong subadditivity, so the splitting on node $k$ is a splitting on the coarse-grained graph.

\begin{figure}
\includegraphics[width=5cm]{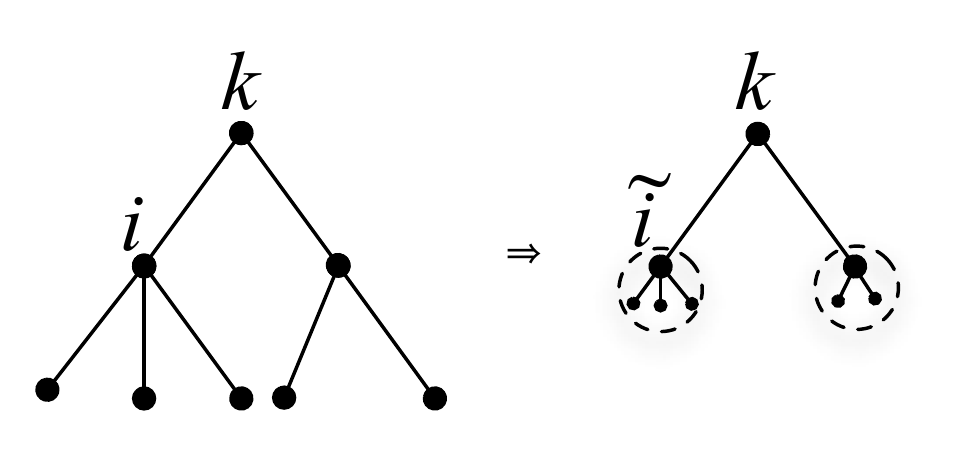}
\caption{Coarse graining procedure.}
\label{fig:Coarse}
\end{figure}

Let $P_k(j)$ project onto $\cH_k(j)$.  Consider any sequence of integers $j_1,...,j_N$.
Define
\be
P(j_1,...,j_N)=\Tr(P_1(j_1) ... P_N(j_N) \rho)
\ee
and
\be
\rho(j_1,...,j_N)=\frac{1}{P(j_1,...,j_N)} P_1(j_1) ... P_N(j_N) \rho P_N(j_N) ... P_1(j_1).
\ee
The state $\rho(j_1,...,j_N)$ is non-zero only on $\cH_1(j_1) \otimes ... \otimes \cH_N(j_N)$, where we claim that it is equal to a product state [it is a product of states on $\cH_{k \rightarrow p(k)}(j_k) \otimes \cH_{p(k) \rightarrow k}(j_{p(k)})$ over all $k$].  
We will prove this claim inductively, by proving that given any tree of $N$ nodes, such that for each node of the tree we have a splitting
of that node with corresponding projectors $P_i(j_i)$, then the state $\rho(j_1,...,j_N)$ has the given product form.
Assume it is true on any tree of at most $N-1$ nodes.  Assume, without loss of generality, that node $1$ has at least two neighbors (if no node has more than two neighbors, we are at the case $N=2$ which is trivial).
The decomposition $P_1(j_1)$ gives a splitting of node $1$ on the coarse-grained graph.  So, the state $P_1(j_1) \rho P_1(j_1)/\Tr(P_1(j_1) \rho)$ is equal to
a product of states $\bigotimes_{\tilde i} \rho_{\cH_{1 \rightarrow \tilde i},\tilde i}$.  For any $\tilde i$, we consider
a tree given by the nodes in the original tree which are in $\tilde i$ in the coarse-grained tree and by the space $\cH_{1 \rightarrow \tilde i}$, considered as a single node.  This tree has at most $N-1$ nodes.  
The splitting that we had on nodes $2,...,N$ on the original tree provides a splitting on the nodes on the new tree in the natural manner.  To see this, note that for any node $k$, if the state $P_k(j_k) \rho P_k(j_k)$ is a product state on the coarse-grained graph with $k$ as the root, then the state $P_1(j_1) P_k(j_k) \rho P_k(j_k) P_1(j_1)$ is also a product state.  Thus, since we have a splitting on the new tree, the state on the new tree has the product structure, so $\rho(j_1,...,j_N)$ does indeed have
the product structure that we claim.

One may directly verify that
\be
P_1(j_1) ... P_N(j_N) \exp(-\sum_k H_k) P_1(j_1) ... P_N(j_N) \propto \rho(j_1,...,j_N).
\ee
So, it suffices to show that the given Hamiltonian $\sum_k H_k$ produces the correct normalization so that
\be
P (j_1,...,j_N)=\Tr\Bigl(P_1(j_1) ... P_N(j_N) \frac{1}{Z} \exp(-\sum_k H_k)\Bigr).
\ee
However, the probability distribution $P_1(j_1,...,j_N)$ is a classical Markov tree and so by the Hammersley-Clifford theorem the desired
result follows (this result can also be proven inductively in roughly the same way as the previous paragraph).
\end{proof}
\end{theorem}

\end{document}